\documentclass[amsmath,twocolumn,twoside,letterpaper,final,10pt]{iopart}                
\usepackage{amsfonts}
\usepackage{amssymb}
\usepackage{setstack}
\usepackage{mathbbol}
\usepackage{graphicx}
\usepackage{txfonts}
\usepackage{color}
\usepackage{nonfloat}
\usepackage{multicol}
\newcommand{\eqleft}{\hspace*{-2.5cm} }
\newcommand{\newu}[1]{{\color{black}{#1}}}

\newcommand{\old}[1]{}
\usepackage{citesort}
\DeclareMathAlphabet{\mathpzc}{OT1}{pzc}{m}{it}

\newcommand\myfigure[1]{%
\medskip\noindent\begin{minipage}{\columnwidth}
\centering%
#1%
\end{minipage}\medskip}

\begin{document}


\title[Modelling the effect of nuclear motion on the attosecond TRPES of ethylene]{Modelling the effect of nuclear motion on the attosecond time-resolved photoelectron spectra of ethylene} 

\author{A Crawford-Uranga$^{1,*}$, U De Giovannini$^1$, D J Mowbray$^1$, S Kurth$^{1,2}$ and A Rubio$^{1,*}$}
\address{$^1$ Nano-Bio Spectroscopy Group and ETSF Scientific Development Center, Departamento de F\'isica de Materiales, Centro de F\'isica de Materiales CSIC-UPV/EHU-MPC and DIPC, Universidad del Pa\'is Vasco UPV/EHU, Avenida de Tolosa 72, E-20018, San Sebasti{\'a}n, Spain}
\address{$^2$ IKERBASQUE, Basque Foundation for Science, E-48011, Bilbao, Spain}
\eads{\mailto{alisonc1986@gmail.com}, \mailto{angel.rubio@ehu.es}}

\date{\today}

\begin{abstract}
Using time dependent density functional theory (TDDFT) we examine the energy, angular and 
time-resolved photoelectron spectra (TRPES) of ethylene in a pump-probe setup. 
To simulate TRPES we expose ethylene to an ultraviolet (UV) femtosecond pump pulse,
followed by a time delayed extreme ultraviolet (XUV) probe pulse. 
Studying the photoemission spectra as a function of this delay provides us direct access to the dynamic 
evolution of the molecule's electronic levels.  
Further, by including the \old{\newu {nuclear}}\newu{nuclei's} motion, we provide direct chemical insight into the 
chemical reactivity of ethylene. 
These results show how angular and energy resolved TRPES could be used to 
directly probe electron and \old{\newu {nuclear}}\newu{nucleus} dynamics in molecules.


\noindent(Some figures may appear in colour only in the online journal)

\end{abstract}

\small

\noindent{\rule{\textwidth}{0.5pt}}
\begin{multicols}{2}

\footnote[0]{
\begin{minipage}{0.5\columnwidth}
\hspace{-0.1in}$^*$\noindent \footnotesize Authors to whom any correspondence should be addressed.
\end{minipage}
}

\vspace{-3em}
\section{Introduction}

Time-resolved photoelectron spectroscopy (TRPES) is a \newu {well established} technique 
\old {in the short-pulse XUV regime}
for characterizing the electronic and nuclear dynamics occurring after 
photoabsorption in molecules~\cite{trpes1,trpes2,trpes3,trpes4,trpes5,trpes6,trpes7,trpes8,trpes9}.
It allows one to map the occupied electronic states of a given interacting system and complements the 
information one can gather from optical spectroscopy.
TRPES is particularly suited to the study of ultrafast non-adiabatic processes 
because photoelectron spectroscopy is sensitive to both 
electronic configurations and \newu {nuclear} dynamics. 

Many open questions remain, linked to how composite electron-\old{\newu{nuclear}}\newu{nucleus} excitations 
or shake-up processes appear in time-resolved spectroscopies~\cite{Pazourek:2012cw}.
Much work has been done in solid state physics 
concerning phonon side-bands in photoelectron spectra~\cite{Smallwood:2012jr,Graf:2011jw},
and in molecular systems using both standard optical spectroscopies and time resolved transient absorption spectroscopy~\cite{zewail2000,foggi2001}.
With the advent of attosecond laser pulse technology ~\cite{BrabecKrausz:01,Plaja:2013hx}, 
its increased temporal resolution has allowed the direct observation and control of coherent electronic motion~\cite{Sansone:2010ke,Siu:2011ib,Zhou:2012iy}.

In TRPES experiments, a time-delayed probe laser photo-ionizes an electron out to an evolving 
(usually laser-generated) excited state. The outgoing electron's kinetic 
energy and angular distribution is then measured as a function of time~\cite{Wu:2011cv}. 

The time-resolved photoelectron angular distribution (TRPAD), 
is frequently available in gas phase experiments. 
However, it is rarely analysed, 
since it is so difficult to interpret. 
This is because the random orientation of molecules 
in free space significantly broadens the results, 
\newu{and it is also complicated to} observe the ejection dynamics in the molecular frame. 
For this reason, molecular pre-alignment~\cite{Bisgaard:2009hr} and coincidence 
techniques~\cite{Gessner:2006cw}, have proven necessary to extract molecular properties from TRPADs.
Even in the molecular frame, a quantitative analysis is a formidable task. 
This is due to factors such as the multi-electron nature of the system, the subtle coupling
between nuclear and electronic degrees of freedom, and  nonlinearities due to the relatively high laser
intensities required.

To complement these advanced experimental techniques, accurate 
and robust theoretical methodologies are necessary 
for their clear interpretation~\cite{6}.
However, the choice of which level of theory to use is often a difficult task.

On the one side, the direct solution of the full time-dependent Schr\"odinger equation for 
interacting electrons should provide a full description. However, in
the presence of time-dependent external fields, the full solution is feasible only for two electrons 
in three dimensions~\cite{Feist_etal:08,IshikawaUeda:12,Argenti:2013ku}. 
On the other side, the single active electron approximation~\cite{Schafer:1990gga}\newu{,} often invoked in many theoretical works, \newu{is likely to break down for large molecules. 
This has been indicated by recent experiments in the strong field regime} \cite{Boguslavskiy:2012fh}. 

\newu{Several approaches have been employed to model TRPES and TRPAD experiments on molecular systems. 
These include methods based on Hartree-Fock theory coupled to the Schwinger equation~\cite{Lucchese:1982fga,trpes10}; 
schemes involving the projection onto states calculated with static density functional theory \cite{SECOND} or multi-scale second order perturbation theory 
with the ab initio multiple spawning method \cite{trpes11,trpes12};
the Wigner distribution approach with on-the-fly dynamics~\cite{FIRST}, and partitioning techniques applied to coupled equation schemes~\cite{Mignolet:2014em}.
Most of these methods evaluate matrix elements involving continuum states, which are difficult to describe.
}

\old{
To date, the most promising ab-initio approaches
within TDDFT are based on space partitioning
schemes that do not require explicit evaluation of
continuum states.

Time-dependent density functional theory (TDDFT)~\cite{ullrichbook,TDDFTbook} 
\newu{provides an attractive alternative for two main reasons: (i) like any
 other explicitly time-dependent approach, this method does not require an
 explicit evaluation of continuum states and (ii) TDDFT provides a formally
 exact (though in practice always approximate ) framework for the dynamics of
 interacting many-electron systems at a reasonable computational cost.}

TDDFT offers a simple framework to describe many-electron systems 
interacting with external electromagnetic fields. The simplicity resides in the mapping of the 
complicated many-body problem onto a problem of non-interacting electrons in the presence of an 
effective time-dependent single-particle potential. The theory is formally exact and in 
principle can give access to the exact time-dependent density of the 
interacting system \cite{RungeGross:84}. In practice, however, one not only has to make 
an approximation for the unknown exchange-correlation (xc) functional, but also for the specific 
density functional of the physical observable under consideration.}

\newu{Time-dependent density functional theory (TDDFT)~\cite{ullrichbook,TDDFTbook} combined with a space partitioning scheme provides an attractive alternative. 
This is because this method does not require an explicit evaluation of continuum states.}
The sampling-point method has been successfully used to study the PES of small clusters~\cite{PohlReinhardSuraud:00,PohlReinhardSuraud:04}.  
However, this method has suffered from numerical limitations in both angular resolution and total pulse length. 
The recently introduced mask-method~\cite{UMBERTO}, provides both better performance and full momentum resolution.
\newu{Furthermore, it is non-perturbative and includes any interference between different 
ionization channels. } 
It is this method which we employ in the present paper.

Previously  we demonstrated how TDDFT can be used to describe time-resolved spectroscopies \cite{UMBERTO1}. 
In principle, this allows one to handle large scale systems at a reasonable computational cost.  
However, we only considered the electronic degrees of freedom without analysing the impact of nuclear vibrations. 

In the present work we extend the previous 
theoretical TDDFT framework by including classical molecular dynamics (MD) through an Ehrenfest approach~\cite{Alonso:2008fz,Andrade:2009tu}.
We illustrate the combined framework for the case study of ethylene already studied in the literature \cite{trpes11,trpes12}.  
In particular, we investigate the effect of the nuclear degrees of freedom on the time evolution of the electronic $\pi \to \pi^*$ 
transition.

\section{Methodology}
 
Within TDDFT all physical properties of a system
can be determined by knowing their functional dependence with respect to the 
interacting many-body density~\cite{RungeGross:84}. 
The crucial idea of both DFT and TDDFT, is to obtain this many-body 
density through a mapping from the density of a fictitious, auxiliary system of non-interacting 
electrons. The latter is the so-called Kohn-Sham (KS) system~\cite{KS}.   

First, the ground state density is obtained by solving the KS equations self-consistently at the DFT level. 
The evolution of the system then follows according to the Time Dependent Kohn Sham (TDKS) equations
(in atomic units):

\begin{eqnarray}\label{eq:TDKS}
  \eqleft i\frac{\partial \varphi_i(\mathbf{r},t)}{\partial t} &=& 
  \left(-\frac{1}{2}\nabla^2
+ V_{KS}[n](\mathbf{r},t)\right) \varphi_i(\mathbf{r},t),
\end{eqnarray}
for $i = 1,\ldots, N/2$, where
\begin{equation} 
\eqleft  n(\mathbf{r},t) =\sum_{i=1}^{N/2} 2|\varphi_i(\mathbf{r},t)|^2.
\end{equation}
Here we assume an even number of electrons $N$, and a spin-restricted configuration in which all 
the KS spatial orbitals $\varphi_i$ are doubly occupied. The non-interacting electrons move in the KS potential $V_{KS}$ defined as:
\begin{eqnarray}
  \eqleft V_{KS}[n](\mathbf{r},t)&&=\newu{V_{las}(\mathbf{r},t) + V_{ne}(\mathbf{r},t)}\nonumber\\
\eqleft &&+\int d \mathbf{r}\, \frac{n(\mathbf{r^\prime},t)}{|r-r^\prime|} + V_{xc}[n](\mathbf{r},t),
\end{eqnarray}
where \newu{$V_{las}(\mathbf{r},t)= \old{rE}\newu{\mathbf{r}\cdot\mathbf{E}}(t) - \sum_j \old{R}\newu{\mathbf{R}}_j(t)\old{E}\newu{\cdot\mathbf{E}}(t)$ 
is the potential describing the pump and probe laser fields where $\mathbf{E}(t)$
is the electric field associated to the laser, 
$V_{ne}(\mathbf{r},t)$ is the electron-\old{nuclear}\newu{nucleus} potential}, 
the \newu{third} term is the Hartree potential and 
$V_{xc}[n](\mathbf{r},t)$ is the exchange and correlation potential.   
However, the exact form for the exchange and correlation potential is generally unknown.

The electron-nucleus potential, for a system composed of $M$ atoms, is given by:
\begin{equation}
  \eqleft \newu{V_{ne}(\mathbf{r},t)} = -\sum_{j=1}^{M} \frac{Z_j}{|\mathbf{R}_j\newu{(t)}-\mathbf{r}|},
\end{equation}
where \newu{$\mathbf{R}_j(t)=\{\mathbf{R}_1(t),\ldots, \mathbf{R}_M(t)\}$} are
the set of classical \old{nuclear}\newu{nucleus} positions and $Z_j$ are their respective charges.
The electronic density depends parametrically on the \old{nuclear}\newu{nuclei's} positions \newu{$\mathbf{R}_j(t)$}. 
The motion of the nuclei is in turn determined by the electronic density gradient through 
Newton's equations of motion. 
The motion of nucleus $j$, with mass $M_j$, evolves according to the following dynamic equation:
\begin{eqnarray}\label{eq:classicalForces}
  \eqleft M_j\frac{d^2 \mathbf{R}_j(t)}{dt^2}&& 
  =\int d \mathbf{r}\, n(\mathbf{r},t)\nabla_j \newu{[V_{ne}(\mathbf{r},t) + V_{las}(\mathbf{r},t)]} \nonumber \\
\eqleft  && + \nabla_j \sum_{\ell\neq j}\frac{Z_{\ell} Z_j}{|\mathbf{R}_j(t)-\mathbf{R}_{\ell}(t)|} \,.
\end{eqnarray}
The Ehrenfest MD scheme consists of the time propagation of the coupled equations (\ref{eq:TDKS}) and 
(\ref{eq:classicalForces}).

Photoelectrons are obtained using a space partitioning scheme~\cite{UMBERTO}.
At each time step, every KS orbital 
$\varphi_i(\mathbf{r},t)=\varphi_i^A(\mathbf{r},t)+\varphi_i^B(\mathbf{r},t)$ is divided into the part residing in an inner interaction region  $A$, $\varphi_{i}^A(\mathbf{r},t)$, and
the remainder $\varphi_i^B(\mathbf{r},t)$ 
in the complementary  ionization region $B$.
In region $A$ we solve the coupled TDKS plus classical nuclear motion equations in the presence of a mask boundary absorber of a given width $R_{ab}$. 
Absorbed electrons are collected and 
evolved in momentum space as free Volkov states $\tilde{\varphi}_i^B(\mathbf{p},t)$ within the ionization 
region $B$. 
The boundary between $A$ and $B$ has to be placed at sufficiently large distances so that 
$\tilde{\varphi}_i^B(\mathbf{p},t)$ is composed of outgoing waves only.
The momentum-resolved photoelectron probability is then
\begin{equation}
\eqleft P(\mathbf{p})=\lim_{t\rightarrow\infty}\sum_{i=1}^{N/2} 2|\tilde{\varphi}_i^B(\mathbf{p},t)|^2.
\end{equation}
Observables such as PES $P(E)$ or PAD $P(E,\theta,\phi)$ are obtained from $P(\mathbf{p})$ by 
integration or slicing respectively.

The exchange and correlation functional we use in this work is the well known local density 
approximation (LDA) coupled with an average-density self interaction correction 
(ADSIC)~\cite{Perdew:1981dv,Legrand:2002jf} for the ground state and its adiabatic extension for TDDFT. 
The choice of ADSIC is motivated by its correct asymptotic behaviour in the ground state.  
In other words, for a large distance $r$ from the molecule, $V_{xc} \sim -1/r$. 
The high-lying unoccupied KS bound states close to the ionization threshold are thus described 
more accurately than with an exponentially decaying xc potential.
For ethylene, using LDA alone gives an unbound $\pi_z^*$ state. This makes
the  $\pi_z \to \pi_z^*$ transition inaccesible with most standard xc-functionals.
On the other hand, the combination of LDA and ADSIC has been successfully employed 
in conditions similar to the ones described in this work~\cite{ZhiPing:2010ej,Wang:2011by}.

In addition, we freeze the $1s$ electrons of the carbon cores into a 
pseudopotential generated within the Troullier-Martins scheme~\cite{TM} as 
distributed in the {\sc octopus} code. \newu{In this way we ``dress'' the carbon nucleus.  }As these core levels are strongly 
bound ($\sim$10.29 Ha for carbon) we expect that neither the pump nor the 
probe will ionize them.

The TDKS equations are discretized and solved using a finite differences method
within the {\sc octopus} code \cite{OCTOPUS,OCTOPUS1,OCTOPUS2}. We employ 
a spherical grid of radius $R=30~a_0$ with a grid spacing of $\Delta R = 0.3~a_0$.
We introduce a $10~a_0$ wide mask boundary absorber to collect the photoelectrons and
prevent electronic reflections~\cite{UMBERTO,UMBERTO1}.
A velocity Verlet algorithm is employed to propagate Newton's equations (\ref{eq:classicalForces})
and an enforced time-reversal symmetry operator \cite{Castro:2004hk} is used to time-step the TDKS equations, 
with a time step of $\Delta t= 1.2$ as. This is small enough to steadily propagate 
nuclear and electronic degrees of freedom in all the presented calculations.

The molecular geometry obtained by force minimization ($\leq 2.4
\times 10^{-5}$~$\frac{\mathrm{Ha}}{a_0}$) has the two carbon atoms placed on 
the $x$-axis at $x=\pm 1.169~a_0$ and hydrogens in the $xy$-plane at $(x,y)=(\pm2.120,\pm1.785)~a_0$.
The carbon-carbon (C--C) bond-length of 2.337~$a_0$ is in fair agreement with the experimental one of 
2.531~$a_0$~\cite{C2H4}. 

\myfigure{
\begin{center}
\includegraphics[width=1\columnwidth]{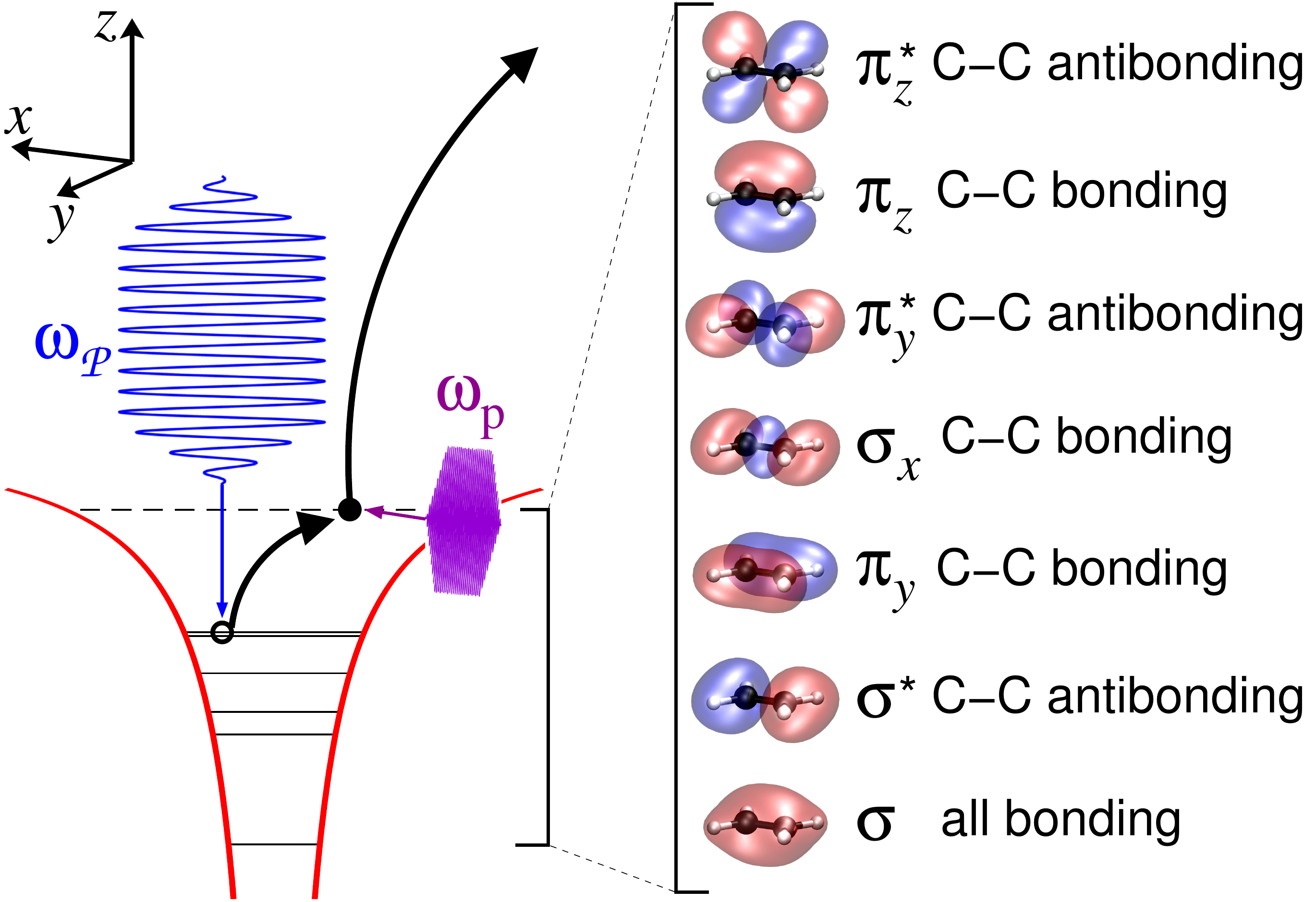}
\end{center}
\figcaption{
Schematic of the pump-probe setup employed to study the TRPES of ethylene. 
The pump (blue) is an ultraviolet (UV) laser pulse of energy 
$\omega_{\mathpzc{P}} = 0.326$~Ha, 
with a 15 cycle trapezoidal shape (3 cycle ramp), 
and an intensity $I = 1.67\times10^{11}$~W/cm$^2$. 
polarized along the C--C bond, i.e., $x$ axis. 
The excited electron is probed using an extreme ultraviolet (XUV) probe laser (violet) 
of energy $\omega_{\mathrm{p}} = 1.8$~Ha, with a
40 cycle trapezoidal shape (8 cycle ramp), and an intensity of $I=1.02\times10^{11}$~W/cm$^2$
polarized along the $z$ axis. 
The calculated Hartree potential (red), KS eigenvalues (black horizontal lines), 
structural schematics (C in black, H in white), and KS orbitals (isosurfaces of $\pm0.05 e/a_0^{3/2}$) 
for ethylene are shown according to their energy. 
}
\label{fig_1:fig}}

Moreover the first ionization potential obtained from LDA+ADSIC is $I_p =0.447$~$a_0$, in agreement with the experimental 
value of 0.386~$a_0$~\cite{EXP}. 
Here, $I_p$ has been evaluated within DFT using the vacuum energy minus that of the 
highest occupied KS orbital, i.e., $I_p\approx E_{\mathit{vac}} - \varepsilon_{\mathrm{HOMO}}$.


The photoionization process we consider is depicted schematically in 
figure~\ref{fig_1:fig}, where the laser parameters have been adapted from Ref.~\cite{UMBERTO1}.
A pump pulse resonantly populates the bound $\pi^{*}_{z}$ state and the build-up is monitored 
at different times with a delayed probe laser.  
This is accomplished by applying an ultraviolet (UV) pump that is tuned 
to excite from the highest occupied molecular orbital (HOMO) to the 
lowest unoccupied molecular orbital (LUMO). For ethylene, this 
corresponds to a $\pi_z \to \pi^*_z$ transition.

In particular, we employ an UV pump laser of energy $\omega_{\mathpzc{P}} = 0.326$~Ha, with a 
15 cycle trapezoidal shape (3 cycle ramp), and an intensity $I = 1.67\times10^{11}$~W/cm$^2$ polarized 
along the $x$-axis. 
The probe is an XUV laser of energy $\omega_{\mathrm{p}} = 1.8$~Ha, with a 40 cycle 
trapezoidal shape (8 cycle ramp), and an intensity of $I=1.02\times10^{11}$~W/cm$^2$
polarized along the $z$-axis. 
  
The time delay between the pump and the probe is measured from the onset 
of the pump to the center of the probe so that negative delays 
correspond to cases where the probe precedes the pump. 
Moreover, we record photoelectrons only during the on-time of the probe pulse.

When we calculate the spectra for classically moving \newu{nuclei}, we introduce an initial temperature of 300~K
under the same pump-probe conditions. In this way, we may assess the impact of the \newu{nuclear} degrees of freedom on the TRPES.
The temperature is introduced in our model by assigning to each nucleus a random initial 
velocity consistent with a Boltzmann distribution at that temperature.

\section{Results and Discussion}

\begin{figure*}
\begin{center}
\includegraphics[width=1.5\columnwidth]{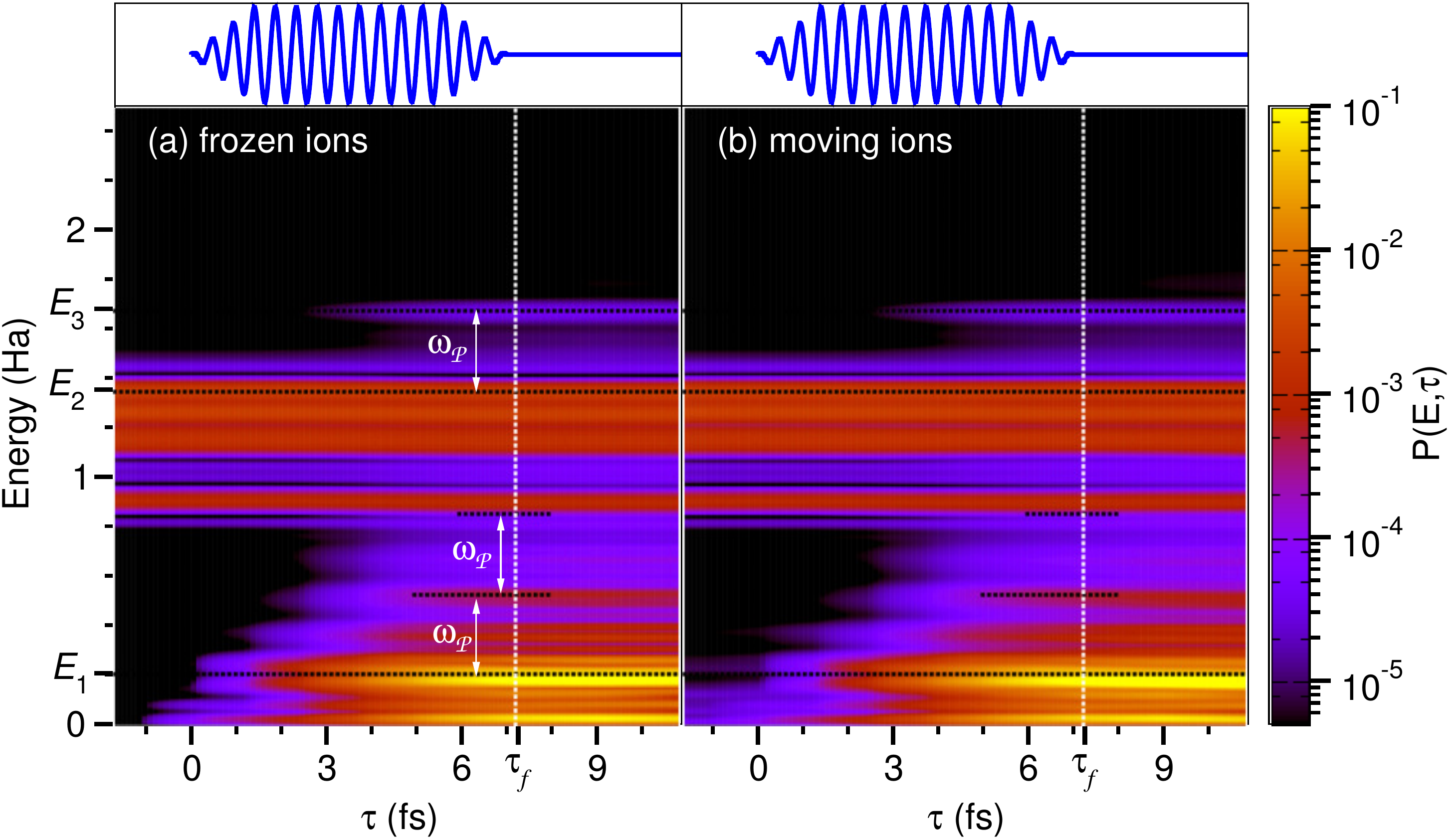}
\end{center}
\figcaption{
TRPES $P(E,\tau)$ of ethylene as a function of the photoelectron's kinetic energy in Ha and the 
pump-probe time delay $\tau$ in fs with frozen (a), or moving (b) \newu{nuclei}. \newu{Nuclear} motion is modelled with an 
initial temperature of 300~K. 
Pump (blue) and probe (violet) pulses are polarized with laser parameters as described in figure~\ref{fig_1:fig}.
The pump is depicted in the upper panels of (a) and (b) as a function of $\tau$.
Here $E_1=2\omega_{\mathpzc{P}} - I_p$, $E_2 = \omega_{\mathrm{p}} - I_p$, $E_3=  \omega_{\mathpzc{P}}+  \omega_{\mathrm{p}}-I_P$.
The energies $E_1, E_1+\omega_{\mathpzc{P}}, E_1+2\omega_{\mathpzc{P}}, E_2, E_3$ (\dotted\ black)
and the time delay $\tau_f$ (\dotted\ white) are shown to guide the eye.
White arrows correspond to the pump's energy $\omega_{\mathpzc{P}}$.
}
\label{fig_2:fig}
\end{figure*}

The TRPES for ethylene with the \newu{nuclei} frozen at their equilibrium positions is shown in figure~\ref{fig_2:fig} (a).
The spectra presents similar features to those described in Ref.~\cite{UMBERTO1}.  
 
The peak at $E_1=2\omega_{\mathpzc{P}} - I_p = 0.205$~Ha constitutes the main ionization 
channel and it is due to the pump alone.
Here, the absorption of a pump photon leading to a $\pi_z \to \pi^*_z$ transition, is followed by a second pump photon 
which directly excites electrons from the $\pi^*_z$ state into the continuum. 
Multi-photon replicas of this peak can be observed at energies separated by integer multiples of
$\omega_{\mathpzc{P}}$.

The peak at $E_2 = \omega_{\mathrm{p}} - I_p =1.353$~Ha corresponds
to the direct emission from the highest occupied KS orbital into the continuum.

A similar mechanism, but with electrons ejected from deeper levels, is responsible 
for the peaks lying at energies lower than $E_2$. These peaks depend on molecular ground state
properties and the probe laser only. For this reason they can be observed also for negative delays
$\tau < 0$.

The population of the $\pi_{z}^{*}$ state increases with the delay for $\tau>0$.
At about the same $\tau$ for which $E_1$ becomes visible, the peak at 
$E_3=\omega_{\mathpzc{P}}+\omega_{\mathrm{p}}-I_P = 1.679$~Ha begins to emerge. 
This peak corresponds to electrons ejected into the continuum 
from the $\pi^{*}_z$ state,
which is transiently occupied via the pump pulse.

To further analyse the results, in figure~\ref{fig_3:fig} 
we plot a cut of the TRPES at $\tau_f=7.26$~fs, 
after the pump has been switched off.
In the figure we introduce new peak labels in addition to $E_1$, $E_2$, $E_3$, 
which have been previously discussed.
These labels identify the contribution to each peak 
from the ground state KS orbitals of ethylene shown
in figure~\ref{fig_1:fig}.

\myfigure{
\begin{center}
\includegraphics[width=1.\columnwidth]{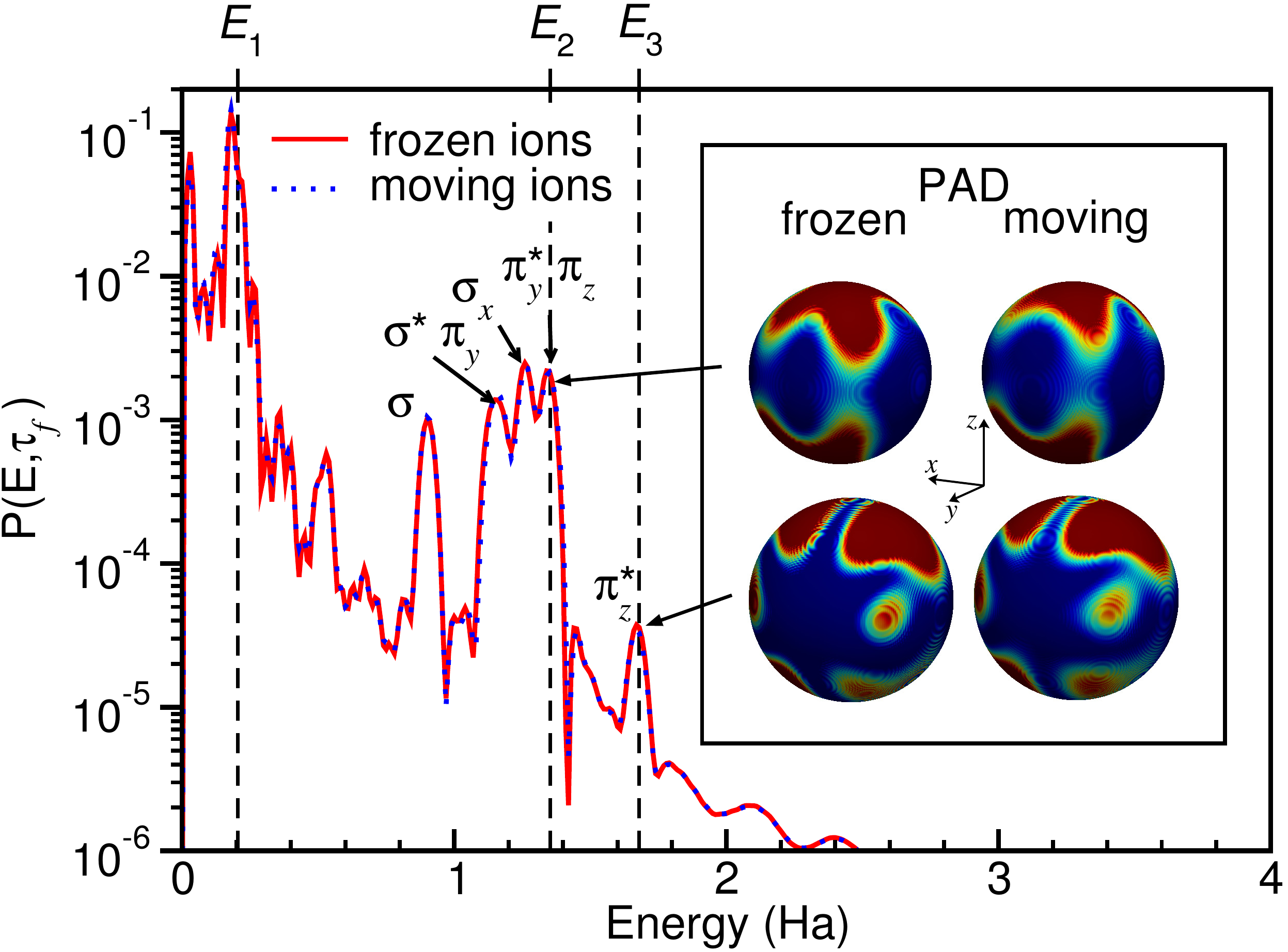}
\end{center}
\figcaption{Photoemission spectra of ethylene versus the photoelectron's kinetic energy in Ha 
for a probe applied at the end of the pump ($\tau_f = 7.26$~fs as shown in figure~\ref{fig_2:fig}) $P(E,\tau_f)$.  
The \newu{nuclei} are either frozen (red), or their motion is classically modelled at an initial temperature of 300 K (blue). 
Peaks at $E_1, E_2$ and $E_3$ correspond to the energy transitions described in figure~\ref{fig_2:fig}, 
while those labelled $\sigma$, $\sigma^*$, $\pi_y$, $\sigma_x$ and $\pi_y^*$ correspond 
to direct excitations by the probe from the respective orbitals depicted in figure~\ref{fig_1:fig}. 
PADs for ethylene at the energies $E_2$ and $E_3$ when the pump has ended $\tau_f$ for frozen and moving \newu{nuclei}
are shown in the inset.
}
\label{fig_3:fig}}

\newu {Supplementary information regarding the nature of PES peaks can be obtained from the PADs.
For electrons ejected from orbitals with $\pi$-symmetry, the momentum resolved photoelectron probability is 
approximately $P(\mathbf{p})\sim |\mathbf{A}\cdot \mathbf{p}|^2 |\tilde{\varphi}(\mathbf{p})|^2$, 
where $\tilde{\varphi}(\mathbf{p})$ is the Fourier transform of the initial orbital~\cite{Puschnig:2009ho} 
and $\mathbf{A}(t)=\int_0^t d\tau \mathbf{E}(\tau)$ is the vector potential in the velocity gauge.
PADs for angles close to the laser polarization direction, where the polarization factor $|\mathbf{A}\cdot 
\mathbf{p}|$ is close to unity, reflect the nodal symmetry of the orbital from which the electron
has been ejected.}

The inset of figure~\ref{fig_3:fig} shows how the 
frozen and moving \old{\newu{nuclear}}\newu{nuclei's} PADs correlate with the originating orbital symmetry.
Photoelectrons emerging with kinetic energy $E_2$ are ejected from almost degenerate $\pi^*_y$ and $\pi_z$
orbitals. The PADs associated with $E_2$ coherently display a symmetry compatible with the superposition 
of these orbitals. On the other hand, the PADs for $E_3$ present a nodal structure clearly linked to a $\pi_z^*$ orbital
symmetry.


As shown in figure~\ref{fig_2:fig}(b), the effect of \newu{nuclear} motion on the electronic TRPES is negligible.
Further, the PAD is minimally changed by the effect of moving the \newu{nuclei} as shown in the inset of 
figure~\ref{fig_3:fig}.

\newu{On the one hand, the occupation of the anti-bonding $\pi^*_z$ orbital is not significant throughout the simulation. 
There are two main reasons for this outcome.
The first one traces back to known problems in describing resonant state population with adiabatic TDDFT. 
This is related to inaccuracies in the time dependence 
of the xc-kernel~\cite{Fuks:2013fn}. The second reason is the depopulation of the $\pi^*_z$ orbital through the 
direct ionization channel observed at $E_1$.}

\myfigure{
\begin{center}
\includegraphics[width=\columnwidth]{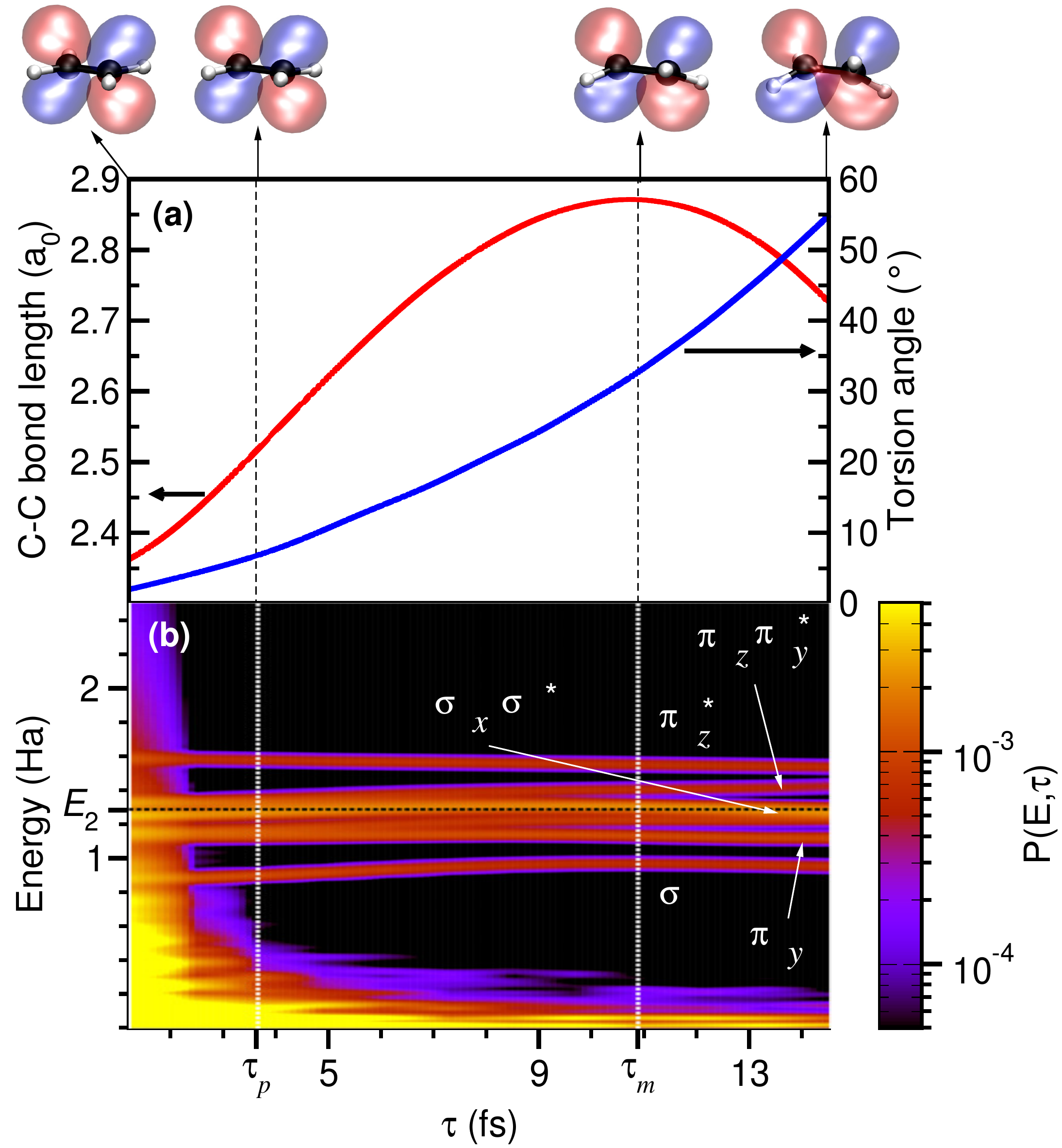}
\end{center}
\figcaption{
(a) Ethylene carbon-carbon bond length in~$a_0$ (red) and torsion angle in~$^\circ$ (blue) 
versus the probe time delay $\tau$ in fs from an artificial $\pi_z \to \pi_z^*$ excited initial state.
The \old{\newu{nuclear}}\newu{nuclei's} motion is classically modelled starting from 300 K and the ground state. 
The black vertical lines are the times at which the C--C 
bond length reaches its maximum $\tau_m = 10.8$~fs
and a previous time $\tau_p = 3.6$~fs for comparison.
The molecular structure and $\pi_z^*$ orbital at the start, $\tau_p$, $\tau_m$ and end of the simulation
are shown above.
(b) Probe TRPES $P(E,\tau)$ \newu{for all the labelled orbitals of ethylene in figure~\ref{fig_1:fig},}  
as a function of the photoelectron's kinetic energy in Ha and the 
probe time delay $\tau$ in fs (see figure~\ref{fig_1:fig} for details of the probe) 
starting from the ground state. 
The \newu{nuclei}'s motion is classically modelled with an initial temperature of 300 K.  
The black \dotted\ vertical lines are the times at which the C--C 
bond length reaches its maximum $\tau_m = 10.8$~fs
and a previous time $\tau_p = 3.6$~fs for comparison.
}
\label{fig_4:fig}}

\newu{On the other hand, changes of the TRPES due to nuclear motion are also small because they are calculated over a large energy range (3Ha).  This is done to include 
the evolution of all the ethylene orbitals shown in figure~\ref{fig_1:fig}}.

\old{due to \newu{two main reasons. First, the TRPES
is calculated for a large energy range (3Ha), to include 
the evolution of all the ethylene orbitals shown in figure~\ref{fig_1:fig}.
For this energy range, the nuclear motion is indistinguishable.
Second, the time scale of the applied pump and probe lasers, is quite
long compared to the time scale of nuclear motion.}
The direct ionization channel observed at $E_1$
depopulates the anti-bonding $\pi_z^*$ state.}



As a result, the molecular 
geometry is minimally modified during the action of the pump pulse, with a maximum change 
in the C--C bond-length of less than 0.033~$a_0$. 
For the laser parameters depicted in figure~\ref{fig_1:fig}, 
the photoelectron properties of the molecule are largely unaffected 
by the coupling with \newu{nuclear} degrees of freedom.
\newu{ 
Although the bandwidth of the probe (~$\sim 0.05$~Ha)
can resolve electronic transitions, it is too wide to resolve vibrational and thermal effects.
As a result, effects due to the initial temperature do not appear in the TRPES~\cite{WOLF}.}

\myfigure{
\begin{center}
\includegraphics[width=\columnwidth]{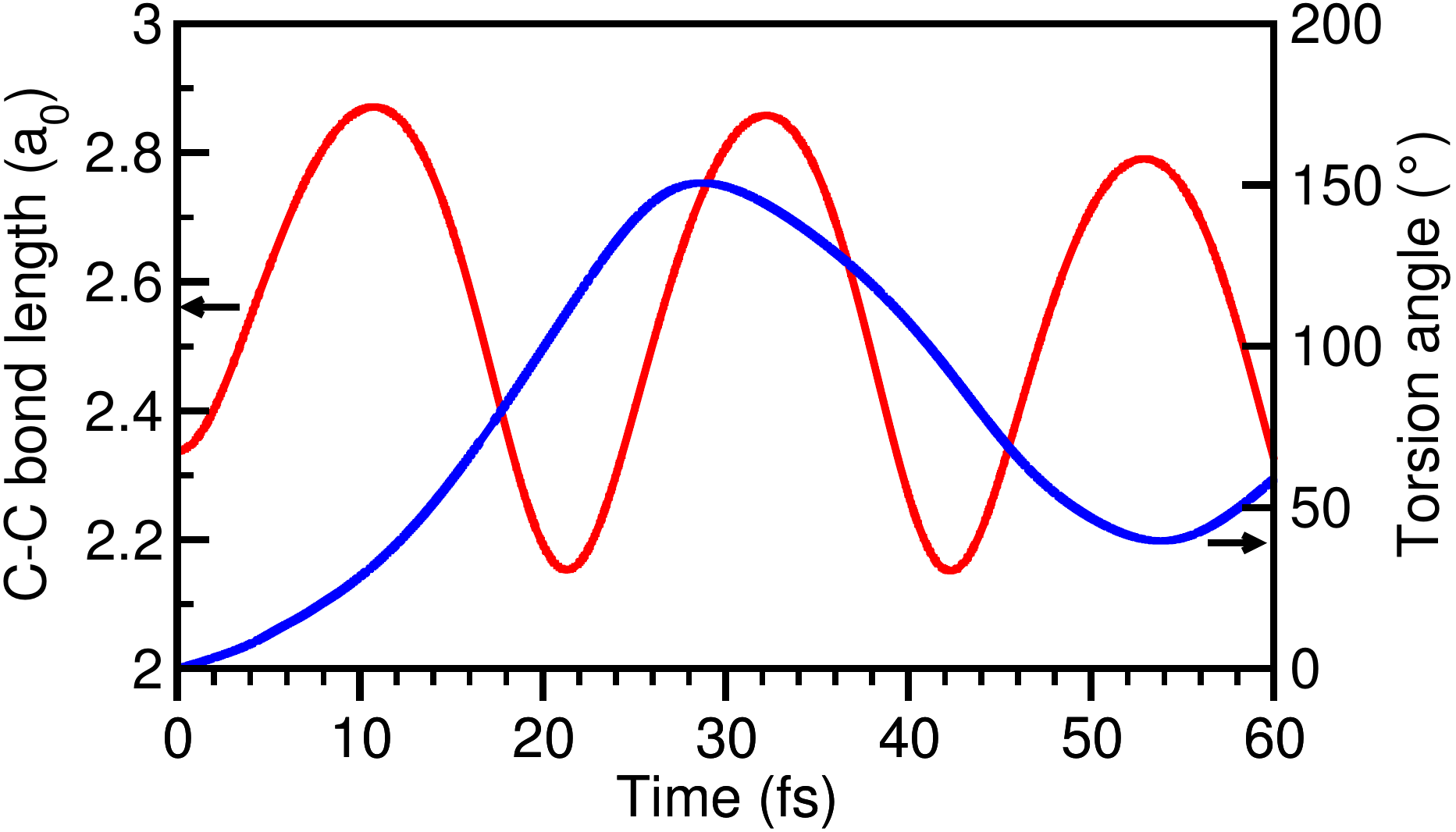}
\end{center}
\figcaption{
Ethylene carbon-carbon bond length in~$a_0$ (red) and torsion angle in~$^\circ$ (blue) 
as a function of time in fs, 
with the \old{\newu{nuclear}}\newu{nuclei's} motion classically modelled starting from 300 K and 
a molecular excited state created by artificially promoting one electron from the KS HOMO to the LUMO.
}
\label{fig_5:fig}}

A stronger \newu{nuclear} response can be stimulated by propagating a fully occupied electronic excited state.
In the previous case, the pump laser was in charge of populating an excited state, 
which was subsequently observed during its construction, by means of a delayed probe pulse. 
We now investigate the effect of the coupling between \newu{nuclear} and 
electronic degrees of freedom while keeping an excited state fully populated.

To this end, we artificially promote one electron from the highest occupied KS
molecular orbital (HOMO) to the first unoccupied one (LUMO), and propagate keeping this configuration. 
The KS LUMO with $\pi^*$ symmetry is of anti-bonding nature. We thus expect its
occupation to have sizeable effects on the \old{\newu{nuclear}}\newu{nuclei's} motion, especially on the C--C bond. 
We employ the same probe pulse shown in figure~\ref{fig_1:fig}, while the pump pulse has been omitted.
Here, the time delay is measured as the difference from the center of the probe to the starting point of the 
time evolution.

Changes in the bond length and the torsion of the molecule 
induced by the initial electronic excitation are shown 
on the left and right hand side of figure~\ref{fig_4:fig}~(a), respectively. 
The same is shown in figure~\ref{fig_5:fig} for a longer time propagation.
The C--C bond length displays an oscillatory behaviour. It initially
increases up to $0.53$~$a_0$ at $\tau_m = 10.8$~fs over its initial 
ground state equilibrium position, and then oscillates in time.

The molecule undergoes a twist along the C--C axis
reaching a maximum torsion of $150.6^\circ$.
This behavior is at the core of cis-trans isomerization processes happening in 
many photochemical reactions~\cite{Kunert:2005cy}.
The vibrational stretching frequency along the C--C bond ($\omega_{CC}=7.14 \times 10^{-3} $~Ha)
and the torsional distortion 
\myfigure{
\begin{center}
\includegraphics[width=\columnwidth]{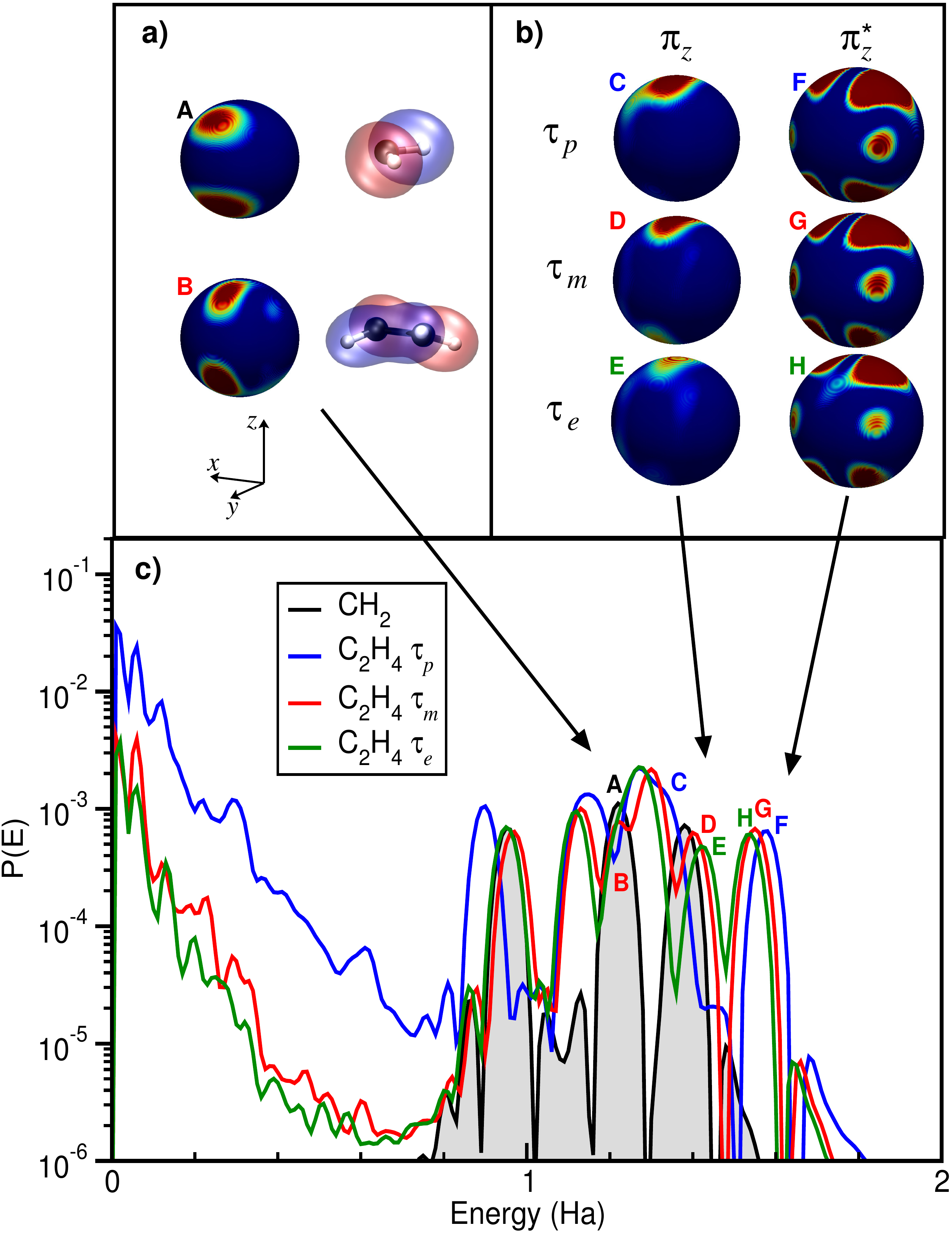}
\end{center}
\figcaption{
(a) PADs and $\pi_y$ orbitals for methylene (A) and ethylene (B).
(b) Ethylene PADs for $\pi_z/\pi_z^*$ orbitals at $\tau_p,\tau_m$ and $\tau_e$.
(c) PES for a probe applied to the methylene molecule and 
to the ethylene molecule for the time at which the C--C bond length 
reaches its maximum $\tau_m = 10.8$~fs, an earlier time $\tau_p = 3.6 < \tau_m$~fs 
and at the end of the simulation $\tau_e = 14.5$~fs.
The PAD for the peaks labelled in the PES for both ethylene and methylene are shown in (a) and (b).
}
\label{fig_6:fig}
}
($\omega_{torsion}=2.82 \times 10^{-3} $~Ha)
are in qualitative agreement with the experimental data
($\omega_{CC}= 7.39\times 10^{-3}$~Ha and $\omega_{torsion}= 4.67\times 10^{-3}$~Ha)~\cite{Kunert:2005cy}.

These modifications of the molecule's geometry are reflected in the TRPES shown in figure~\ref{fig_4:fig}~(b). 
For time delays $\tau \le \tau_p$ we have five initial peaks in the TRPES, as we had in the pump-probe case
(see figures~\ref{fig_2:fig} and~\ref{fig_3:fig}).
However, now the initial spectrum changes in time, as the peaks shift in position and new ones emerge close 
to the maximum elongation of the C--C bond.

The lowest energy peak oscillates in energy in phase with the C--C bond length.
This peak is consistent with all-bonding $\sigma$ orbital electrons (see figure~\ref{fig_1:fig}), and 
is therefore sensitive to the molecule's bond-length and relatively insensitive to its torsion angle.

The following peak in energy splits into two new peaks. 
The peak that shifts in energy, corresponds to the $\pi_y$ state,
whereas the one that does not, to the $\sigma^*$ state.
The $\pi_y$ state energy shift is due to the fact that it connects hydrogens bound to different carbon atoms, 
and is therefore sensitive to the molecule's torsion and the C--C bond length.

The intermediate peak in energy corresponds to the $\sigma_x$ state, which does not shift in energy, 
as it is only weakly affected by the molecule's bond stretching.


The second to last peak in energy corresponds to the $\pi_y^*$ and $\pi_z$ states. 
The former shifts towards lower energies because it does not connect hydrogens bound to different carbon atoms,
and the latter depletes probability along the nodal plane, shifting towards higher energies.

The highest and last peak in energy is consistent with the $\pi_z^*$ and it shifts towards lower energies
because it builds probability along the nodal plane. 

The $\pi_z^*$ and $\pi_z$ orbitals become degenerate when a torsion angle of $90^\circ$ is reached.
The LUMO $\pi_z^*$ orbital evolution is shown on top of figure~\ref{fig_4:fig}.

In order to support this analysis, in figure~\ref{fig_6:fig}~(c) we present 
selected cuts of the TRPES at the specific time delays $\tau_p= 3.6$~fs, $\tau_m=10.8$~fs and 
$\tau_e=14.5$~fs. The time evolution of each photoelectron peak, 
can here be monitored identifying each peak with its PAD {\em fingerprint} in figures~\ref{fig_6:fig}~(a) and (b). 
The peaks label\newu{l}ed \textbf{F}, \textbf{G}, \textbf{H}, which shift towards lower energies as time evolves, 
belong to the same state according to the PADs. This state can easily be associated to a $\pi_z^*$ orbital due to its nodal structure.
Similarly, the peaks \textbf{C}, \textbf{D}, \textbf{E}, which shift towards higher energies as time evolves, 
all originate from the same $\pi_z$ orbital. 

In comparison to the HOMO $E_2$ PADs we observed  
in the pump-probe case of figure~\ref{fig_3:fig}, the 
$\pi_z$ character is here more defined. 
This is because the occupation of the KS LUMO state
is lifting the $\pi_y^*$, $\pi_z$ degeneracy that we had previously. 
When the C--C elongation is at its maximum value at $\tau_m$, a new peak 
emerges at $E_{\rm \mathbf{B}}=1.22$~Ha, which disappears at $\tau_e$.

In order to understand where this extra peak comes from, 
we have obtained the photoelectron spectrum of methylene with the same probe pulse
used for ethylene as shown filled in figure~\ref{fig_6:fig}~(c).
The PADs and orbitals of the peaks label\newu{l}ed \textbf{B} (belonging to ethylene) 
and \textbf{A} (belonging to methylene) in figure~\ref{fig_6:fig}~(a) 
display a $\pi_y$ symmetry. We can therefore conclude, 
that this extra peak is related to the $\pi_y$ ethylene state 
which becomes less stable as the C--C bond length increases.

The first main peak corresponds to a $\sigma$
state, which increases in energy until $\tau_m$ and then decreases again until $\tau_e$ as shown in
the PES in figure~\ref{fig_6:fig}~(c). The second main peak shifts towards lower energies
as time evolves. This corresponds to the $\sigma^*$ and $\pi_y$ states, which separate in energy at $\tau_m$, as explained
above.

\section{Conclusions}

In this paper we investigated the impact of \newu{nuclear} degrees of freedom in TRPES and TRPAD
for the test case of the ethylene molecule.

We first studied the case where a pump laser resonantly excites a bound state which 
is subsequently probed by a time delayed pulse with attosecond time scale resolution.
\newu{The applied pump does not induce a sufficient occupation of the anti-bonding $\pi^*_z$ orbital
for changes in the nuclear positions to be resolved by the probe. 
The photoelectron spectrum is therefore minimally modified on the energy scales considered.}

In order to induce detectable \newu{nuclear} effects we studied the time evolution of a molecular 
excited state created by artificially promoting one electron from the KS HOMO to the LUMO.
This \newu{promotion} has proven to be sufficient to excite vibrational C--C bond and torsional modes. 
TRPES in this case has shown \newu{major} changes that can be understood in terms of PADs and
orbital deformations associated to \newu{nuclear} rearrangements.

\newu{This work is an initial step towards a scalable TDDFT scheme for ab-initio simulation of time resolved photoemission processes 
coupled with nuclear motion. Further effort will be spent in the future on the road to the inclusion of nuclear effects beyond Ehrenfest.}

\ack
We acknowledge financial support from the European Research Council Advanced Grant
DYNamo (ERC-2010-AdG-267374), Spanish Grants (FIS2010-21282-C02-01 and PIB2010US-00652),
Grupos Consolidados UPV/EHU del Gobierno Vasco (IT-578-13) and Ikerbasque. A. Crawford-Uranga
also acknowledges financial support from the Departamento de Educaci\'on, Pol\'tica Lingu\'istica
y Cultura del Gobierno Vasco (BFI-2011-26) and DIPC. D. J. Mowbray
acknowledges financial support from the Spanish ``Juan de la Cierva'' program 
(JCI-2010-08156).

\section*{References}


\begin{thebibliography}{10}
\expandafter\ifx\csname url\endcsname\relax
  \def\url#1{{\tt #1}}\fi
\expandafter\ifx\csname urlprefix\endcsname\relax\def\urlprefix{URL }\fi
\providecommand{\eprint}[2][]{\url{#2}}

\bibitem{trpes1}
Haight R 1995 {\em Surf. Sci. Rep.\/} {\bf 21} 275

\bibitem{trpes2}
Cyr D~R and Hayden C~C 1996 {\em J. Chem. Phys.\/} {\bf 104}

\bibitem{trpes3}
Greenblatt B~J, Zanni M~T and Neumark D~M 1997 {\em Science\/} {\bf 276} 1675

\bibitem{trpes4}
Petek H and Ogawa S 1997 {\em Prog. Surf. Sci.\/} {\bf 56} 239

\bibitem{trpes5}
Blanchet V, Zgierski M~Z, Seideman T and Stolow A 1999 {\em Nature (London)\/}
  {\bf 401} 52

\bibitem{trpes6}
Lehr L, Zanni M~T, Frischkorn C, Weinkauf R and Neumark D~M 1999 {\em
  Science\/} {\bf 284} 635

\bibitem{trpes7}
Neumark D~M 2001 {\em Annu. Rev. Phys. Chem.\/} {\bf 52} 255

\bibitem{trpes8}
Reid K~L 2003 {\em Annu. Rev. Phys. Chem.\/} {\bf 54} 397

\bibitem{trpes9}
Stolow A 2003 {\em Annu. Rev. Phys. Chem.\/} {\bf 54} 89

\bibitem{Pazourek:2012cw}
Pazourek R, Feist J, Nagele S and Burgd{\"o}rfer J 2012 {\em Phys. Rev.
  Lett.\/} {\bf 108} 163001

\bibitem{Smallwood:2012jr}
Smallwood C~L, Hinton J~P, Jozwiak C, Zhang W, Koralek J~D, Eisaki H, Lee D~H,
  Orenstein J and Lanzara A 2012 {\em Science\/} {\bf 336} 1137--1139

\bibitem{Graf:2011jw}
Graf J, Jozwiak C, Smallwood C~L, Eisaki H, Kaindl R~A, Lee D~H and Lanzara A
  2011 {\em Nat Phys.\/} {\bf 7} 805--809

\bibitem{zewail2000}
Zewail A 2000 {\em J. Phys. Chem. A\/} {\bf 104} 5660

\bibitem{foggi2001}
Foggi P, Bussotti L and V~R~Neuwahl F 2001 {\em International Journal of
  Photoenergy\/}  103--109

\bibitem{BrabecKrausz:01}
Brabec T and Krausz F 2000 {\em Rev. Mod. Phys.\/} {\bf 72} 545--591

\bibitem{Plaja:2013hx}
Plaja L, Torres R and Za{\"\i}r A (eds) 2013 {\em {Attosecond Physics}\/} ({\em
  Springer Series in Optical Sciences\/} vol 177) (Berlin, Heidelberg: Springer
  Berlin Heidelberg)

\bibitem{Sansone:2010ke}
Sansone G, Kelkensberg F, Perez-Torres J~F, Morales F, Kling M~F, Siu W, Ghafur
  O, Johnsson P, Swoboda M, Benedetti E, Ferrari F, Lepine F, Sanz-Vicario J~L,
  Zherebtsov S, Znakovskaya I, LHuillier A, Ivanov M, Nisoli M, Martin F and
  Vrakking M~J~J 2010 {\em Nature\/} {\bf 465} 763--766

\bibitem{Siu:2011ib}
Siu W, Kelkensberg F, Gademann G, Rouzee A, Johnsson P, Dowek D, Lucchini M,
  Calegari F, De~Giovannini U, Rubio A, Lucchese R, Kono H, Lepine F and
  Vrakking M~J 2011 {\em Phys. Rev. A\/} {\bf 84}

\bibitem{Zhou:2012iy}
Zhou X, Ranitovic P, Hogle C~W, Eland J~H~D, Kapteyn H~C and Murnane M~M 2012
  {\em Nat Phys.\/} {\bf 8} 232--237

\bibitem{Wu:2011cv}
Wu G, Hockett P and Stolow A 2011 {\em Phys. Chem. Chem. Phys.\/} {\bf 13}
  18447--18467

\bibitem{Bisgaard:2009hr}
Bisgaard C~Z, Clarkin O~J, Wu G, Lee A~M~D, Gessner O, Hayden C~C and Stolow A
  2009 {\em Science\/} {\bf 323} 1464--1468

\bibitem{Gessner:2006cw}
Gessner O, Lee A~M~D, Shaffer J~P, Reisler H, Levchenko S~V, Krylov A~I,
  Underwood J~G, Shi H, East A~L~L, Wardlaw D~M, Chrysostom E~t~H, Hayden C~C
  and Stolow A 2006 {\em Science\/} {\bf 311} 219--222

\bibitem{6}
Freericks J~K, Krishnamurthy H~R and Pruschke T 2009 {\em Phys. Rev. Lett.\/}
  {\bf 102} 136401--136405

\bibitem{Feist_etal:08}
Feist J, Nagele S, Pazourek R, Persson E, Schneider B, Collins L and
  Burgd{\"o}rfer J 2008 {\em Phys. Rev. A\/} {\bf 77} 043420

\bibitem{IshikawaUeda:12}
Ishikawa K~L and Ueda K 2012 {\em Phys. Rev. Lett.\/} {\bf 108} 033003

\bibitem{Argenti:2013ku}
Argenti L, Pazourek R, Feist J, Nagele S, Liertzer M, Persson E, Burgd{\"o}rfer
  J and Lindroth E 2013 {\em Phys. Rev. A\/} {\bf 87} 053405

\bibitem{Schafer:1990gga}
Schafer K and Kulander K 1990 {\em Physical Review A\/} {\bf 42} 5794--5797

\bibitem{Boguslavskiy:2012fh}
Boguslavskiy A~E, Mikosch J, Gijsbertsen A, Spanner M, Patchkovskii S, Gador N,
  Vrakking M~J~J and Stolow A 2012 {\em Science\/} {\bf 335} 1336--1340

\bibitem{Lucchese:1982fga}
Lucchese R, Raseev G and McKoy V 1982 {\em Phys. Rev. A\/} {\bf 25} 2572--2587

\bibitem{trpes10}
Arasaki Y, Takatsuka K, Wang K and McKoy V 2010 {\em J. Chem. Phys.\/} {\bf
  132} 124307

\bibitem{SECOND}
Suzuki Y, Stener M and Seideman T 2003 {\em J. Chem. Phys.\/} {\bf 118} 4432

\bibitem{trpes11}
Tao H, Allison T~K, Wright T~W, Stooke A~M, Khurmi C, van Tilborg J, Liu Y,
  Falcone R~W, Belkacem A,  and Martinez T~J 2011 {\em J. Chem. Phys.\/} {\bf
  134} 244306

\bibitem{trpes12}
Tao H, Levine B~G and Martinez T~J 2009 {\em J. Phys. Chem. A\/} {\bf 113}
  13656

\bibitem{FIRST}
Stanzel J, Burmeister F, Neeb M, Eberhardt W, Mitric R, Burgeland C and
  Bonacic-Koutecky V 2007 {\em J. Chem. Phys.\/} {\bf 127} 164312

\bibitem{Mignolet:2014em}
Mignolet B, Levine R~D and Remacle F 2014 {\em Phys. Rev. A\/} {\bf 89} 021403

\bibitem{ullrichbook}
Ullrich C~A 2012 {\em Time-Dependent Density-Functional Theory: Concepts and
  Applications\/} (Oxford University Press)

\bibitem{TDDFTbook}
Marques M, Maitra N, Nogueira F, Gross E and Rubio A (eds) 2012 {\em
  Fundamentals of Time-Dependent Density Functional Theory\/} ({\em Lecture
  Notes in Physics\/} vol 837) (Berlin: Springer)

\bibitem{PohlReinhardSuraud:00}
Pohl A, Reinhard P and Suraud E 2000 {\em Phys. Rev. Lett.\/} {\bf 84} 5090

\bibitem{PohlReinhardSuraud:04}
Pohl A, Reinhard P and Suraud E 2004 {\em Phys. Rev. A\/} {\bf 70} 023202

\bibitem{UMBERTO}
Giovannini U~D, Varsano D, Marques M~A~L, Appel H, Gross E~K~U and Rubio A 2012
  {\em Physical Review A\/} {\bf 85} 062515

\bibitem{UMBERTO1}
Giovannini U~D, Brunetto G, Castro A, Walkenhorst J and Rubio A 2013 {\em
  Chemphyschem\/} {\bf 14} 1363--1376

\bibitem{Alonso:2008fz}
Alonso J, Andrade X, Echenique P, Falceto F, Prada-Gracia D, Rubio A and Rubio
  A 2008 {\em Phys. Rev. Lett.\/} {\bf 101} 096403

\bibitem{Andrade:2009tu}
Andrade X, Castro A, Zueco D, Alonso J~L, Echenique P, Falceto F and Rubio A
  2009 {\em J. Chem. Theory Comput.\/} {\bf 5} 728--742

\bibitem{RungeGross:84}
Runge E and Gross E~K~U 1984 {\em Phys. Rev. Lett.\/} {\bf 52} 997

\bibitem{KS}
Kohn W and Sham L~J 1965 {\em Physical Review\/} {\bf 140} A4: A1133

\bibitem{Perdew:1981dv}
Perdew J~P and Zunger A 1981 {\em Phys Rev B\/} {\bf 23} 5048--5079

\bibitem{Legrand:2002jf}
Legrand C, Suraud E and Reinhard P~G 2002 {\em Journal of Physics B: Atomic,
  Molecular and Optical Physics\/} {\bf 35} 1115--1128

\bibitem{ZhiPing:2010ej}
Zhi-Ping W, Jing W and Feng-Shou Z 2010 {\em Chinese Phys. Lett.\/} {\bf 27}
  013101

\bibitem{Wang:2011by}
Wang Z~P, Dinh P~M, Reinhard P~G, Suraud E and Zhang F~S 2011 {\em Int. J.
  Quantum Chem.\/}  480--486

\bibitem{TM}
Troullier N and Martins J~L 1991 {\em Phys. Rev. B\/} {\bf 43} 1993

\bibitem{OCTOPUS}
Castro A, Marques M, Appel H, Oliveira M, Rozzi C, Andrade X, Lorenzen F, Gross
  E and Rubio A 2006 {\em Phys. stat. sol\/} {\bf 243} 2465--2488

\bibitem{OCTOPUS1}
Andrade X, Alberdi-Rodriguez J, Strubbe D~A, Oliveira M~T, Nogueira F, Castro
  A, Muguerza J, Arruabarrena A, Louie S, Aspuru-Guzik A, Rubio A and Marques M
  2012 {\em J. Phys.: Condens. Matter\/} {\bf 24} 12

\bibitem{OCTOPUS2}
Marques M, Castro A, Bertsch G and Rubio A 2003 {\em Computer Physics
  Communications\/} {\bf 151} 60--78

\bibitem{Castro:2004hk}
Castro A, Marques M~A~L and Rubio A 2004 {\em J Chem. Phys.\/} {\bf 121}
  3425--3433

\bibitem{C2H4}
Craig N~C, Groner P and McKean D~C 2006 {\em J. Phys. Chem. A\/} {\bf 110}
  7461--7469

\bibitem{EXP}
Turner D~W, Baker C, Baker A~D,  and Brundle C~R (eds) 1970 {\em Molecular
  Photoelectron Spectroscopy\/} (London: Wiley)

\bibitem{Puschnig:2009ho}
Puschnig P, Berkebile S, Fleming A~J, Koller G, Emtsev K, Seyller T, Riley J~D,
  Ambrosch-Draxl C, Netzer F~P and Ramsey M~G 2009 {\em Science\/} {\bf 326}
  702--706

\bibitem{Fuks:2013fn}
Fuks J~I, Elliott P, Rubio A and Maitra N~T 2013 {\em J. Phys. Chem. Lett.\/}
  {\bf 4} 735--739

\bibitem{WOLF}
Wollenhaupt M, Engel V and Baumert T 2005 {\em Annu. Rev. Phys. Chem.\/} {\bf
  56} 25--56

\bibitem{Kunert:2005cy}
Kunert T, Grossmann F and Schmidt R 2005 {\em Phys. Rev. A\/} {\bf 72} 023422

\end{thebibliography}
\providecommand{\newblock}{}

\end{multicols}
\end{document}